\documentclass[groupedaddress,reprint,prl]{revtex4-1}
\usepackage{graphicx}
\usepackage{amsmath}

\begin{document}

\title{Customizing optical patterns via feedback-based wavefront shaping}

\author{Peilong Hong}\email{plhong@njust.edu.cn}
\affiliation{School of Optoelectronic Science and Engineering, University of Electronic Science and Technology of China (UESTC), Chengdu 610054, China}

\date{\today}

\begin{abstract}
We show that customized optical patterns can be generated by employing feedback-based wavefront shaping without prior knowledge of the transmission matrix of optical systems.
To control the spatial distribution of intensity within the target region, the overlap coefficient is employed as the key feedback signal for wavefront shaping, and the optimal phase pattern for generating the customized optical pattern is reached when the overlap coefficient converges to unity.
By using the overlap-coefficient optimized wavefront shaping, we generate customized optical patterns such as multiple focused spots of different intensity, cat image, and speckle patterns of tailored intensity distribution.
Besides, we also show that a sharp focusing spot can be generated in highly scattering systems with the overlap-coefficient optimized wavefront shaping, and the achieved signal-to-background ratio is much higher than that achieved with the conventional wavefront shaping.
Our method provides new possibilities to control the interference of light for many applications, such as optical imaging, quantum interference, and optical trapping and manipulation.
\end{abstract}


\maketitle


Customizing optical pattern is essential for many applications, such as structured-illumination imaging~\cite{mudry2012structured,kuplicki2016high}, quantum interference~\cite{defienne2016two,wolterink2016programmable}, secured optical communication~\cite{goorden2014quantum}, and optical trapping and manipulation~\cite{ashkin1986observation}.
Active control of both the amplitude and the phase on the wavefront would be a direct way for generating customized optical pattern, but relies on relatively complicated optical setup with significantly reduced energy efficiency~\cite{goorden2014superpixel,lee1974binary}.
Besides, the phase-only spatial modulator (SLM) that can actively control the phase pattern of wavefront is widely used for efficient generation of the customized optical patterns,
while the optimal phase pattern for generating the customized optical pattern can be obtained through phase-retrieval algorithms~\cite{fienup1982phase,bromberg2014generating,bender2018customizing}.
In general, to obtain the complex or phase-only wavefront for generating a customized optical pattern in an optical system, the transmission matrix from the control plane to the target plane should be known as prior knowledge, which usually needs to be measured in advance due to imperfections or random scattering in real setups.
However, measuring the transmission matrix could be very challenging for complex optical systems, and requires of complicated yet highly stable optical setups, such as the situation with light transmitting through a thick scattering medium~\cite{yu2013measuring,popoff2010measuring}.
In this letter, we show that the optimal phase pattern for generating a customized optical pattern can be retrieved by employing a feedback-based wavefront shaping method.
Since the method is free of measuring the transmission matrix of the optical system, it could be very useful for generating customized optical pattern in many scenarios such as in complex optical systems.

Wavefront shaping (WFS) was pioneered by Vellekoop and Mosk~\cite{vellekoop2007focusing}, in which the intensity measured at the target position is employed as the key feedback signal, i.e., the intensity at target position is optimized. Wavefront shaping has led to many applications in complex scattering systems, such as focusing through a scattering medium~\cite{vellekoop2007focusing,vellekoop2010exploiting}, imaging with a scattering medium~\cite{popoff2010image,katz2012looking,van2011scattering}, enhanced optical transmission through a scattering medium~\cite{vellekoop2008universal,kim2012maximal,popoff2014coherent}, and controlling energy density of light inside a scattering medium~\cite{sarma2016control,hong2018three}. While the conventional feedback-based wavefront shaping can optimize the intensity within the target region, it is not capable of giving an optimal solution for generating a customized optical pattern within the target region~\cite{vellekoop2015feedback}.
To resolve the problem, we propose changing the key feedback signal in wavefront shaping, and instead optimizing the overlap coefficient between the measured intensity pattern and the customized optical pattern. While the constructive interference guarantees the convergence to a global optimum for conventional wavefront shaping~\cite{vellekoop2007focusing,vellekoop2008phase}, the convergence of the overlap-coefficient optimized wavefront shaping (oc-WFS) is less obvious. Next, we first give a detailed description and analysis of the oc-WFS method, and then use it to obtain the optimal phase pattern for generating customized optical patterns in the target region.


\begin{figure}[t]
    \centering
    \includegraphics[width=0.45\textwidth]{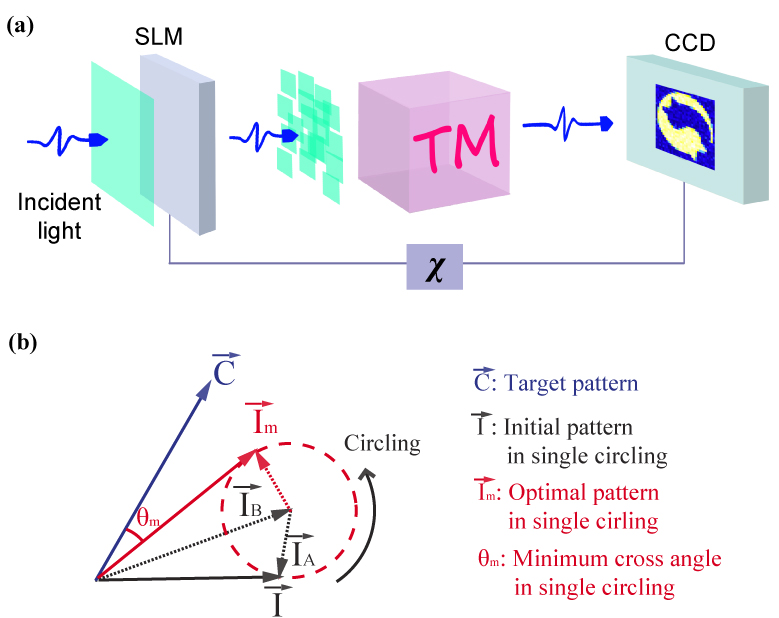}
    \caption{(a): Schematic of generating a customized optical pattern through oc-WFS in a general optical system. The propagation of light from the SLM to the observation plane can be modeled by a transmission matrix (TM) (b): Principle of oc-WFS. An optimal phase for the controlled SLM pixel is obtained in single circling/iteration. 
\label{Figure_1}}
\end{figure}

We consider a general setup as shown in Fig.~\ref{Figure_1}(a).
The incident light is first shaped by a phase-only SLM, and then propagates through the system to reach the observation plane.
When the light transmits through a scattering phase screen or scattering medium located between the SLM and the observation plane, a random speckle pattern will form on the observation plane.
To realize focusing at a target point on the observation plane with conventional WFS, the intensity measured at the target point is employed as the key feedback signal.
By consecutively circling the phase of each pixel of the SLM from 0 to 2$\pi$, the optimal phase that maximizes the feedback signal is stored.
At this optimal phase, constructive interference is built between the field contributed by the controlled pixel and the background field from other pixels.
For the continuous sequential WFS~\cite{vellekoop2008phase}, the obtained optimal phase is loaded on the controlled pixel, and then the process is repeated pixel by pixel such that the complex amplitudes originating different pixels of the SLM interference constructively.
As a result, the shaped field converges to a sharp focusing spot at the target point.
However, when the target region is larger than a single speckle, conventional WFS only optimizes the total intensity within the target region, but is not capable of optimizing a customized optical pattern.

To generate customized optical pattern with WFS, the developed oc-WFS optimizes the overlap coefficient defined as
\begin{equation}\label{ove_coef}
\chi=\frac{\sum_{j=1}^M{C_j \cdot I_j}}{\sqrt{\sum_{j=1}^M{C_j^2}\cdot\sum_{j=1}^M{I_j^2}}},
\end{equation}
where the summation is over all the possible modes within the target regions, and $C_j$ represents the intensity of $j$th mode of the customized optical pattern, while $I_j$ represents the measured intensity of the $j$th mode.
It is clear that $\chi\le 1$, and the customized optical pattern is obtained when $\chi\equiv 1$.
The oc-WFS is of the same procedure as the continuous sequential WFS described above, except that the overlap coefficient is now used as the key feedback signal for optimization.
To understand the convergence of the oc-WFS, we model the customized pattern as vector ${\vec{C}}=(C_1,C_2,\cdots,C_M)$, and the measured intensity pattern as vector ${\vec{I}}=(I_1,I_2,\cdots,I_M)$ in the $M$-dimensional space. 
Consequently, the two vectors are not aligned when $\chi<1$, and is aligned when $\chi\equiv 1$.
Note that each element of the vector ${\vec{I}}$ can be expressed as $I_j=|B_j+A_{jk}|^2=|B_j|^2+|A_{jk}|^2+2|B_j||A_{jk}|cos(\phi_j-\phi_{jk})$, where $A_{jk}$ ($\phi_{jk}$) represents the amplitude (phase) of the field contributed from the $k$th pixel of the SLM, and $B_{j}$ ($\phi_{j}$) represents the amplitude (phase) of the background contributed from the other pixels.
Therefore, ${\vec{I}}$ is the summation of two sub-vectors ${\vec{I}_B}=(|B_1|^2+|A_{1k}|^2,|B_2|^2+|A_{2k}|^2,\cdots,|B_M|^2+|A_{Mk}|^2)$ and  ${\vec{I}_A}=(2|B_1||A_{1k}|cos(\phi_1-\phi_{1k}),2|B_2||A_{2k}|cos(\phi_2-\phi_{2k}),\cdots,2|B_M||A_{Mk}|cos(\phi_M-\phi_{Mk}))$, as conceptually represented in Fig.~\ref{Figure_1}(b).
By circling the phase of $k$th pixel on the SLM, the sub-vector ${\vec{I}_A}$ is circled correspondingly.
The optimal phase that minimizes the cross angle $\theta_m$ between the resulted ${\vec{I}}_m$ and the target ${\vec{C}}$ is stored, and loaded on the $k$th pixel, i.e., $\chi$ is maximized at this optimal phase. 
By continuously repeating the process pixel by pixel, we make the vector ${\vec{I}}$ aligned as close as possible to the vector ${\vec{C}}$ that represents the target optical pattern.
After each pixel of the SLM is addressed for one iteration, a new circling process for the SLM pixels can be started until the overlap coefficient $\chi$ getting close to unity.
Next, we do numerical experiments to show that the oc-WFS is capable of reaching the optimal phase pattern for generating the customized optical patterns.


\begin{figure}[b]
    \centering
    \includegraphics[width=0.45\textwidth]{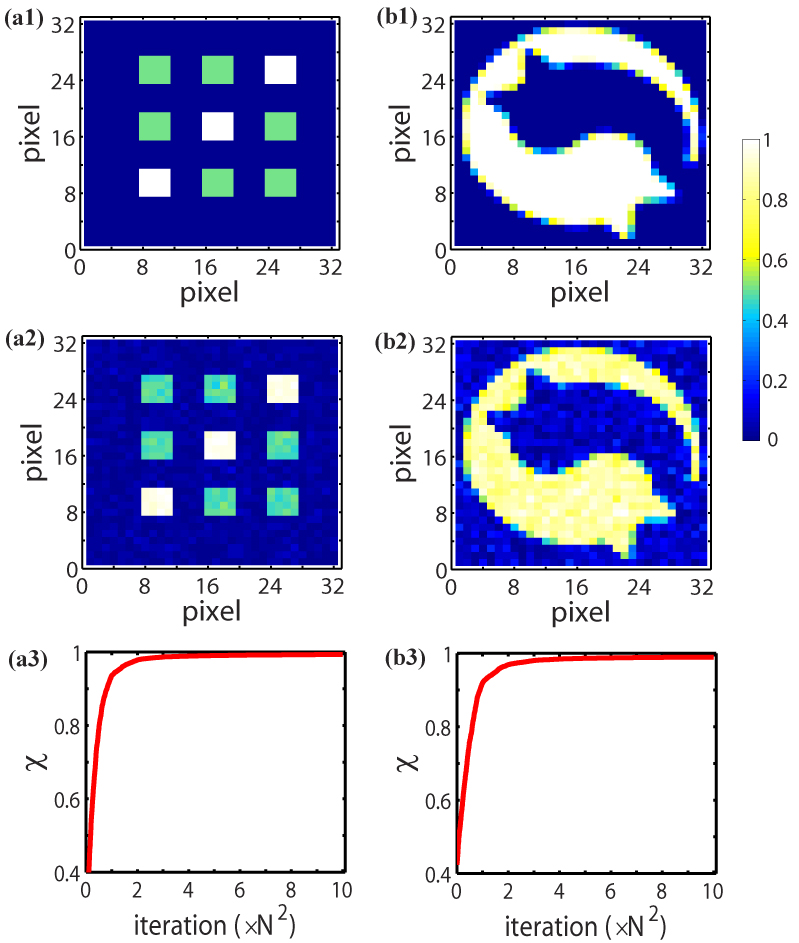}
    \caption{Generating customized optical patterns through oc-WFS.
(a1,b1): Customized optical patterns.
(a2,b2): The generated optical pattern through oc-WFS.
(a3,b3): Convergence curve of $\chi$ v.s. iteration ($N^2=32^2$).
.\label{Figure_2}}
\end{figure}

\begin{figure}[t]
    \centering
    \includegraphics[width=0.45\textwidth]{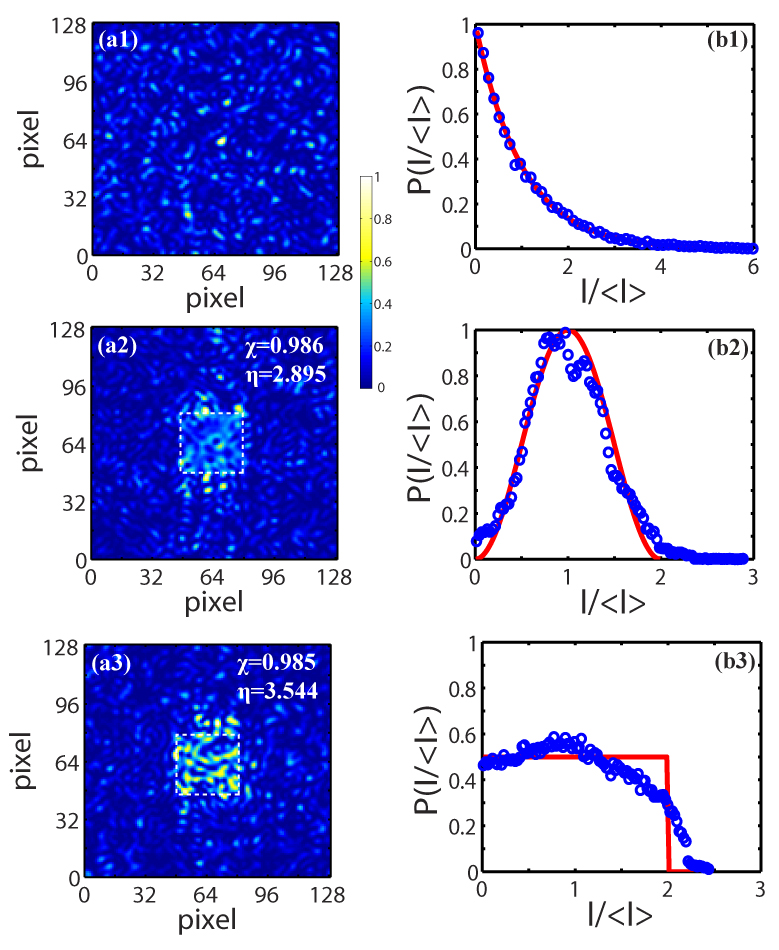}
    \caption{Generating speckle field of tailored distribution through oc-WFS.
(a1,b1): Rayleigh speckle field and probability density distribution of the intensity, respectively.
(a2,b2): The generated optical pattern with unimodal distribution, and the probability density distribution of the intensity, respectively.
(a3,b3):The generated optical pattern with rectangular distribution, and the probability density distribution of intensity, respectively.
Here the target region for oc-WFS is marked by dashed line box. The blue circles in (b1-b3) is the result calculated from the optical image in (a1-a3) respectively, while the red lines in (b1-b3) are theoretical curves.
\label{Figure_4}}
\end{figure}

We first numerically generate two patterns as the customized ones, as shown in Figs.~\ref{Figure_2}(a1) and~\ref{Figure_2}(b1).
Both customized optical patterns are of 32*32 independent pixels.
Without loss of generality, we consider a scheme with the observation plane located at the Fourier plane of the SLM.
For generating the customized optical pattern, the minimum pixel number of the SLM should be 32*32 with each speckle on the Fourier plane representing a corresponding pixel of the customized pattern.
In the simulation, the SLM is initially loaded with a random phase pattern, indicating the method is tolerant to phase perturbations or random scattering.
As a result, the initial optical pattern is a random speckle field on the observation plane, such as that shown in Fig.~\ref{Figure_4}(a1).
To obtain the optimal phase pattern on the SLM plane for generating the customized optical pattern, we numerically implement the oc-WFS procedure, in which the phase is circled from 0 to $2\pi$ by a step size $\pi/8$ for each pixel of the SLM in a iteration.
As the oc-WFS proceeds, the initial speckle pattern on the Fourier plane gradually transforms to the customized one.
With the optimal phase pattern loaded on the SLM, we calculate to obtain the intensity pattern on the observation plane as shown in Figs.~\ref{Figure_2}(a2) and~\ref{Figure_2}(b2), respectively.
One can see that the customized optical patterns have been generated via oc-WFS.
Besides, the convergence behaviors of oc-WFS are shown in Figs.~\ref{Figure_2}(a3) and~\ref{Figure_2}(b3), respectively.
It is seen that the overlap coefficient $\chi$ converges to unity as the number of iterations increases, and considerably high $\chi$ can be achieved within several iterations of the whole SLM.

\begin{figure}[t]
    \centering
    \includegraphics[width=0.45\textwidth]{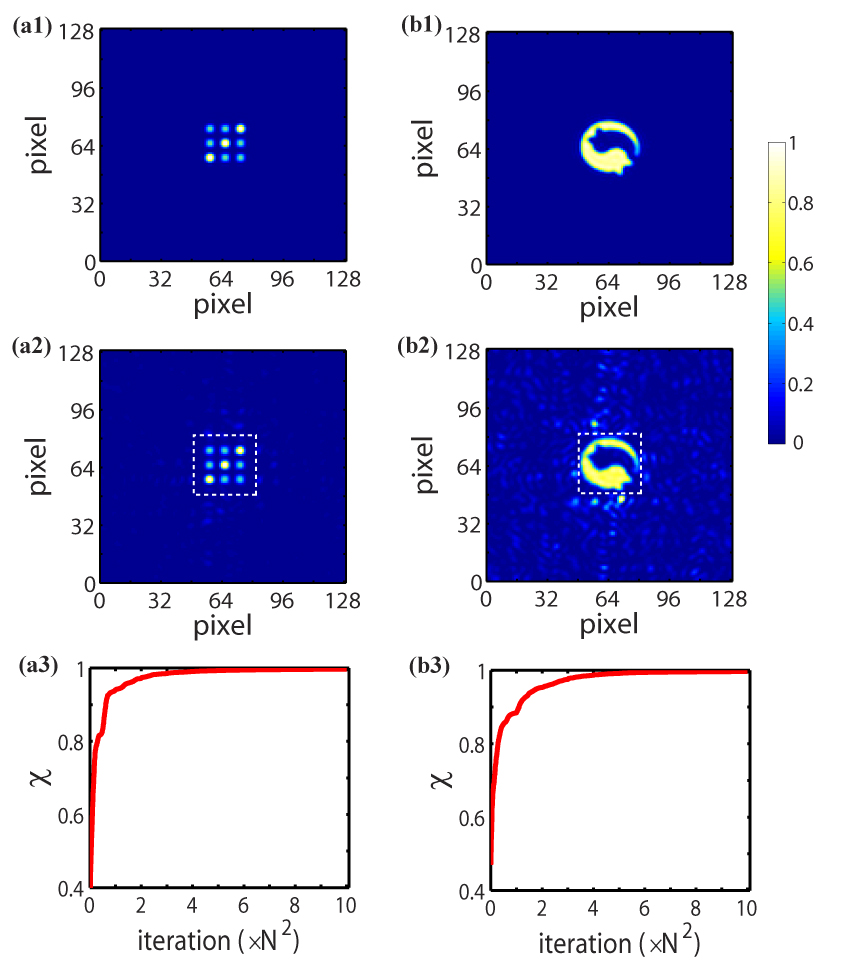}
    \caption{Generating smoothed optical patterns through oc-WFS.
(a1,b1): Smoothed optical patterns.
(a2,b2): The generated optical pattern through oc-WFS.
(a3,b3): Convergence curve of $\chi$ v.s. iteration ($N^2=32^2$).
Here the target region for oc-WFS is marked by dashed line box.
\label{Figure_3}}
\end{figure}

We note that an independent speckle on the Fourier plane is of internal spatial profile, which can be detected by introducing oversampling. Consequently, for generating a customized optical pattern of smooth spatial profile, a speckle in the target region should be sampled at least 2*2 points according to Nyquist limit, i.e., corresponds to at least 2*2 pixels of the customized optical pattern.
To simulate this oversampling situation, we introduce 128*128 sample points on the Fourier plane with each speckle sampled by 4*4 pixels, and then take the 32*32 points in the center of the Fourier plane as the target region for generating a smooth customized optical pattern.
Due to the limited bandwidth related to the speckle size, the customized pattern should be smoothed by filtering out the high frequency part that is outside the range accessible by the optical system.
To realize it, we first Fourier transform the customized pattern to its Fourier space, and filter out the high frequency part that can not be reached by the optical system.
The filtered Fourier image is then transformed back to the real space as the smoothed customized pattern, as shown in Figs.~\ref{Figure_3}(a1) and~\ref{Figure_3}(b1).
In this situation, the generated patterns are shown in Figs.~\ref{Figure_3}(a2) and ~\ref{Figure_3}(b2), while the convergence of overlap coefficient $\chi$ is shown in Figs.~\ref{Figure_3}(a3) and~\ref{Figure_3}(b3), respectively.
One can see that the smooth customized optical pattern is successfully generated in the target region.

To further demonstrate the capability of oc-WFS for generating customized optical patterns, we employ oc-WFS for generating speckle field of tailored distribution.
The target speckle pattern is numerically generated via the same procedure as described in Ref.~\cite{bender2018customizing}.
Again, we employ 128*128 sampling points on the Fourier plane for customizing smooth speckle field in the simulation.
For comparison, the well-known Rayleigh speckle is generated on the Fourier plane by loading a random phase pattern on the SLM.
The obtained Rayleigh pattern is shown in Fig.~\ref{Figure_4}(a1), while the probability density distribution of the normalized intensity $I/\langle I \rangle$ is shown in Fig.~\ref{Figure_4}(b1).
First, we process oc-WFS for generating a customized speckle pattern with unimodal distribution $P(x)=\sin (\pi/2 \cdot x)$ with $x\in [0,2]$.
The obtained pattern through oc-WFS is shown in Fig.~\ref{Figure_4}(a2), while the intensity distribution calculated from the image within the target region is shown in Fig.~\ref{Figure_4}(b2).
It is seen that the target speckle pattern of customized distribution is successfully generated through oc-WFS.
Besides, we also process oc-WFS to generate a speckle pattern with rectangular distribution $P(x)= \text{rect} [(x-1)/2]$, and the results is shown in Figs.~\ref{Figure_4}(a3) and~\ref{Figure_4}(b3), confirming the oc-WFS capable of generating a customized speckle field.
From the results shown in Figs.~\ref{Figure_4}(a2) and~\ref{Figure_4}(a3), we see that the average intensity within the target region is higher than that of the background, although the initial speckle field before processing oc-WFS is of the same average intensity across the whole plane.
We have calculated the ratio $\eta$ of the average intensity within the target region to that of the whole image, it reaches $2.895$ for the generated speckle field with unimodal distribution, and $3.544$ for that with rectangular distribution.
This result shows that the energy efficiency in the oversampling situation can be partially optimized by using oc-WFS.

\begin{figure}[t]
    \centering
    \includegraphics[width=0.45\textwidth]{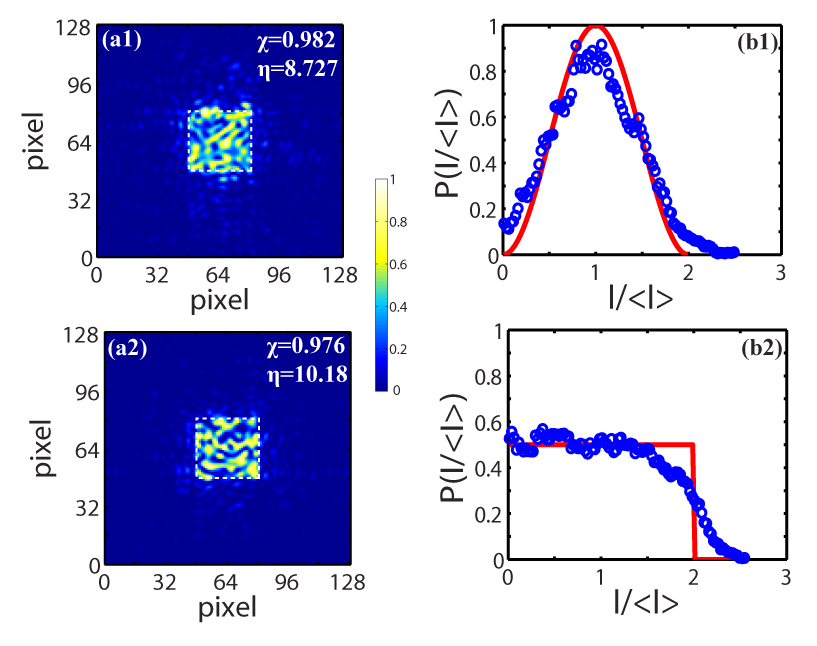}
    \caption{Generating speckle field of customized distribution through oc-WFS with regularization.
(a1,b1): The generated optical pattern with unimodal distribution, and the probability density distribution of the intensity, respectively.
(a2,b2):The generated optical pattern with rectangular distribution, and the probability density distribution of intensity, respectively.
The overlap coefficient $\chi$ and the average intensity enhanced factor $\eta$ are shown in the insets of (a1) and (a2).
\label{Figure_5}}
\end{figure}

Besides, we note that optimizing $\chi$ gives rise to the optimal solution for generating the customized optical pattern, but the total power within the target region is not controlled in the oversampling situation.
To gain partial control of the total energy within the target region, we introduce a regularization term into the feedback signal $\chi$, i.e., optimizing the feedback signal $(\chi + \beta \cdot P_{j}/P_{0})$.
Here $P_{j}$ represents the total power wihtin the target region measured during oc-WFS, while we take the total power $P_{0}$   of the initial speckle field wihtin the target region as the normalization constant.
$\beta$ is a empirical constant, which should be chosen by considering the trade-off between the optimization of overlap coefficient $\chi$ and that of energy efficiency.
With the regularized feedback signal ($\beta=0.01$), we find that the average intensity within the target region is greatly enhanced while the overlap coefficient $\chi$ is still a high value, as shown in Fig.~\ref{Figure_5}.

\begin{figure}[h]
    \centering
    \includegraphics[width=0.48\textwidth]{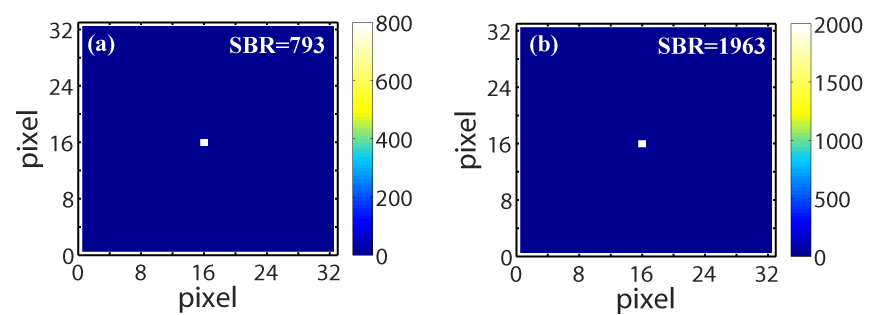}
    \caption{Focusing image realized by conventional WFS (a), and that realized by oc-WFS (b). The intensity in the image is normalized by the average intensity of the background. The calculated SBR for both methods are shown in the inset of the sub-figures, respectively.
\label{Figure_6}}
\end{figure}

The developed oc-WFS is applicable in other complex optical systems, such as that with light transmitting through a multi-mode fiber or a thick scattering medium.
As for example, we did simulation with a scheme by employing transmission matrix to model the propagation of light from the SLM to the observation plane.
The transmission matrix $T$ is of size $N^2 \times N^2$ (N=32), and is generated with the matrix element following the complex Gaussian distribution.
In this case, we set the customized optical pattern as a bright spot at the center while keeping the amplitude elsewhere as zero, i.e., realizing focusing on the observation plane.
As for comparison, the conventional WFS is also introduced for realizing focusing on the observation plane.
The optimized focused image with conventional WFS is shown in Fig.~\ref{Figure_6}(a), while that with oc-WFS is shown in Fig.~\ref{Figure_6}(b).
We calculate the signal-to-background ratio (SBR) that is defined as the ratio of optimized spot intensity to the average background intensity around the optimized spot.
Remarkably, it is found that, comparing with the enhanced SBR ($\sim800$) realized with conventional WFS, the SBR of the optimized image through oc-WFS gains another $\sim1000$ fold enhancement, reaching $\sim1900$. 
To further confirm the result, we have done 10 independent simulations with each method, and the averaged SBR for conventional WFS is $807\pm27$, while that for oc-WFS is $1893\pm104$.
This significant difference may originate from the fact that the conventional WFS only optimizes the focusing spot, while the oc-WFS optimizes both the focusing spot and the background.

In conclusion, we show that the customized optical patterns can be generated in a optical system by employing feedback-based WFS, which does not need the prior knowledge of the transmission matrix.
To obtain the optimal phase pattern for generating the customized optical pattern, the oc-WFS is developed by employing the overlap coefficient for optimization.
The developed oc-WFS converges to a optimal solution that generates the customized optical patterns.
The feedback-based approach can update the optical pattern in real time, which could be of advantage for applications with dynamic perturbation~\cite{vellekoop2015feedback}.
Customizing optical pattern with oc-WFS is useful in many applications, such as image transfer through a scattering medium~\cite{popoff2010image}, constructing multi-port beam splitter with customized transmission for quantum interference~\cite{defienne2016two,wolterink2016programmable}, and controlling the spatial energy density inside a scattering medium~~\cite{sarma2016control,hong2018three}.
The oc-WFS could also be useful for customizing patterns with other waves, such as microwave, x-rays, ultrasonic wave, and electron wave.

\vspace{12pt}
The author thanks the support from the NSFC (Grant No. 11604150).
The author is grateful to Y. Wang for drawing the artwork in Figure~\ref{Figure_1}.

\vspace{12pt}


\end{document}